\documentclass[english,aps,pre,preprint,superscriptaddress,showpacs]{revtex4-1}

\usepackage[T1]{fontenc}
\usepackage[latin9]{inputenc}
\usepackage{float}
\usepackage{amsmath}
\usepackage{amssymb}
\usepackage{graphicx}

\newcommand{\dd}{{\rm d}}

\makeatletter

\newcommand{\noun}[1]{\textsc{#1}}

\usepackage{mathrsfs}
\usepackage[english]{babel}

\makeatother

\usepackage{babel}
\begin{document}

\title{A convergent kinetic equation for gravitational and Coulomb systems.}

\author{Y.~Chaffi}
\affiliation{Physique des systèmes dynamiques. Faculté des Sciences. Université
Libre de Bruxelles (ULB). 1050 Brussels, Belgium.}

\author{T.~M.~Rocha Filho}
\affiliation{Instituto de Fí{}sica and International Center for Condensed
Matter Physics.\\
 Universidade de Brasí{}lia, 70910-900 - Brasí{}lia,
Brazil.}

\author{L.~Brenig}
\email{lbrenig@ulb.ac.be}
\affiliation{Physique des systèmes dynamiques. Faculté des Sciences. Université
Libre de Bruxelles (ULB). 1050 Brussels, Belgium.}

\selectlanguage{english}

\begin{abstract}
It is well known that due to its divergence at large impact parameters,
the Boltzmann collision integral in the kinetic equation for 3D systems of particles interacting
through a $1/r$ potential must be replaced by a Balescu-Lenard-like
collision term. However, the latter diverges at small impact parameters.
This comes from the fact that only weak interactions are considered 
while strong collisions between close particles are neglected in its derivation.
We show that a solution to this dilemma exists in the
framework of the BBGKY formulation of statistical mechanics. It is based on a
separate treatment of the contribution of the strong interactions from that of the weak interactions.
The strong interaction part leads to a new term that involves 
a fractional Laplacian operator in velocity space while the weak interaction component
yields the Balescu-Lenard collision term with an explained lower
cut-off at the Landau length. For spatially uniform initial conditions, the fractional Laplacian
contribution leads to a long-tailed velocity distribution as long as the spatial inhomogeneity remains small. 
We present results from molecular dynamics simulations confirming the existence of such long tails. 
\end{abstract}
\maketitle

\section{Introduction}
\label{sc1}

The validity of the Boltzmann kinetic equation for systems of particles
interacting via the 2-body gravitational or Coulomb potential is a
long-standing problem~\cite{key-10-1}. The long range of the potential
in $\frac{1}{r}$ makes it difficult to reconcile them with the main
condition on which the Boltzmann equation is based, that is, that
binary collisions must be well separated events in space and time.
This condition is realized for diluted systems with short-range interactions.
However, interactions in self-gravitating and Coulomb systems, obviously,
are not short-ranged. As a consequence, during a given binary collision
in such systems, the two particles involved in it are subjected to
many weak interactions exerted on them by the rest of the particles
in the system. Binary collisions, hence, are not well separated events
in these systems. The cumulative effect of the many long distance
weak interactions due to all the other particles on any couple of
colliding particles causes the divergence at large impact parameters
of the integral of the differential cross-section appearing in the
Boltzmann collision term.

A first adaptation of the Boltzmann kinetic equation to systems with
binary interaction potential in $1/r$ has been made by L.D.Landau~\cite{key-24}.
In his derivation, he assumed that all the binary interactions in
the system are weak. Hence, during a binary interaction the velocity
increment of the particles must be small. This assumption allowed
L.D.Landau to expand the integrand in the Boltzmann collision term
into a series in powers of the velocity increment and to keep the
dominant term. This led to the important Landau kinetic equation,
a central tool for describing the collisional regime in plasmas. However,
this equation suffers from two divergences. One of them is inherited
from the above described divergence of the Boltzmann collision integral
at large impact parameters. However, the Landau collision integral
diverges also in the limit of small impact parameters. This is due
to the weak coupling assumption that implies the neglect of the strong
collisions. Both divergences are remedied by introducing cut-offs
that are justified by phenomenology.

In order to cure the divergence at large impact parameters, R.Balescu~\cite{key-11}
and A.Lenard~\cite{key-18} separately derived for weakly coupled
plasmas a kinetic equation that includes the collective effects of
the numerous weak long-range interactions acting on any couple of particles engaged
in a binary collision. Their approaches led to a description of the
interactions in term of a self-consistent screened interaction potential
that is short-ranged but diverges at the origin as $1/r$. The resulting
kinetic equation fulfills the objective of the authors and is free
of the divergence at the large impact parameter limit. The extension
of this theory to self-gravitating systems, however, is not straightforward.
Indeed, such systems are intrinsically inhomogeneous while the Balescu-Lenard
approach crucially depends on the hypothesis of local homogeneity.
For plasmas this property is guaranteed by the screening of the interaction
potential which makes it effectively short-ranged. Nevertheless,
J.Heyvaerts~\cite{key-19} and P.-H.Chavanis~\cite{key-20} managed
to derive a Balescu-Lenard-like for inhomogeneous gravitational systems.

However, like in the Landau approach, these Balescu-Lenard-like theories
rest upon the condition that all the binary interactions are weak
in the system. In doing so, they omit the strong interactions between
very close particles that are, in contrast, well described in the Boltzmann equation.
Though rare, these events cause a divergence of the Balescu-Lenard-type
collision integral in the limit of small impact parameter. This divergence
is the question addressed in the present article.

To summarize, for systems governed by a two-body interaction potential
in $1/r$, the Landau collision term is an approximation of the Boltzmann
term and it diverges at large impact parameters in the same way as
the latter. At the opposite limit, the Boltzmann collision integral
converges for vanishing impact parameter, while the Balescu-Lenard
term diverges in the same limit. However, the latter converges in
the large impact limit. Moreover, the Landau collision term arises
as an approximation from the Balescu-Lenard term when the collective
effects are neglected~\cite{key-10-2} and it conserves the short distance 
divergence of the latter. So the Landau term diverges at both short and long distances.
This observation led some authors ~\cite{key-21}~\cite{key-18-1}~\cite{key-23} to the
intuition that by adding the Boltzmann and Balescu-Lenard collision terms
and subtracting the Landau collision term one would compensate the
divergences at both limits and produce a convergent collision term.
These works proceed from the quantum description and after some approximations
take the classical limit. However, the justification of these approximations remains 
rather unclear and dictated by the expected result.

We show in this work that a derivation of a convergent
kinetic equation is possible from the first principles of classical statistical
mechanics. It consists essentially in separating the contribution of the strong 
interactions between close particles from that of the weak interactions between distant particles. 
At first glance, the strong interactions component should look as a part of the Boltzmann collision term.
However, in contrast with the latter, the conditions for markovianization in time are not
fulfilled. This leads to a new term that involves a fractional time integral and a fractionary 
power of the Laplacian operator in velocity space. The weak interaction part gives, as expected, 
a Balescu-Lenard collision term with a justified cut-off at small distances.

The article is structured as follows. In the second chapter we derive the new kinetic equation. 
Then, in the third chapter,
we consider the kinetic equation for spatially uniform systems and
for times much shorter than the relaxation time associated to the
Balescu-Lenard collision term. This equation is linear and can be
exactly solved yielding a velocity distribution with an algebraic tail in ${1}/{v^{5/2}}$. 
Chapter~\ref{sc4} presents simulations
in molecular dynamics that confirm the short-time existence of this
long-tailed velocity distribution. In chapter~\ref{sc5}, we discuss an apparent
difficulty arising form the divergence of the second moment of the
long-tailed velocity distribution for uniform systems. We solve it and show that
the conserved quantities are well-defined for the new kinetic equation.
Results and perspectives are discussed in the last chapter. Detailed
calculations are exposed in the Appendices.

\section{Deriving the kinetic equation from the BBGKY hierarchy}
\label{sc2}
We now derive the kinetic equation. More details are given in Appendix~\ref{appendixa}.
We consider a system of $N$ identical classical point-like particles
of mass $m$ in $\mathbb{R}^{3}$. They interact via a binary potential
$U(\vec{r})=\gamma/r$. The variable $r$ is the norm of the distance
vector $\vec{r}$ between the two interacting particles. In order
to cover both repulsive and attractive interactions in charged particles
gases and gravitational systems, the coupling constant $\gamma$ can
be positive or negative. 
The statistical state of the system is given by the $N$-particle distribution function
$f_N(\vec{r}_1,\ldots\vec{r}_N,\vec{v}_1,\ldots\vec{v}_N;t)$ given the probability density in phase space for a
state of the $N$-particle system, and obeying the Liouville equation:
\begin{equation}
\frac{\partial f_N}{\partial t}+\left\{f_N,H\right\}=0,
\label{liouvilleeq}
\end{equation}
where $H$ is the Hamiltonian and $\{f,H\}$ the Poisson bracket of $f_N$ and $H$.
For a system of identical particles we assume that $f_N$ is fully symmetric for particle permutations.
The $s$-particle reduced distribution function is defined by (here we adopt the normalization used in~\cite{key-10-2}):
\begin{equation}
f_s({\bf 1},\ldots,{\bf s};t)=\frac{N!}{(N-s)!}\int \dd{\bf (s+1)}\cdots \dd N\:f_N({\bf 1},\ldots,{\bf N};t).
\label{spartred}
\end{equation}
In Eq.~(\ref{spartred}) we used the notation ${\bf 1}\equiv (\vec{r}_1,\vec{v}_1)$, $\dd{\bf 1}\equiv d\vec{r}_1d\vec{v}_1$
and so on for each particle index. By applying the integral operator involved in Eq.~(\ref{spartred}) to the Liouville equation~(\ref{liouvilleeq})
we obtain after some algebra the BBGKY hierarchy for the reduced distribution functions~\cite{key-10-2}):
\begin{equation}
\frac{\partial}{\partial t} f_s=\sum_{j=1}^s\hat{L}_j^0 f_s+\sum_{j<k=1}^s\hat{L}^\prime_{jk}f_s
+\sum_{j=1}^s\int \dd({\bf s+1})\:\hat{L}^\prime_{j,s+1}f_{s+1},
\label{bbgky2}
\end{equation}
where the free motion operator is defined as
$$
\hat L_{i}^{0}\equiv-\vec{v}_{i}\cdot\frac{\partial}{\partial\vec{r_{i}}},
$$
the interaction operator by:
$$
\hat L_{ij}'\equiv-\frac{1}{m}\vec{F}(\vec{r_{i}}-\vec{r}_{j})\cdot(\frac{\partial}{\partial\vec{v}_{i}}-\frac{\partial}{\partial\vec{v}_{j}}),
$$
and
$$
\vec{F}(\vec{r_{i}}-\vec{r}_{j})\equiv\vec{F}_{ij}\equiv\gamma\frac{\vec r_i-\vec r_j}{\left|\left|\vec r_i-\vec r_j\right|\right|^3},
$$
is the interparticle force of particle $j$ on particle $i$. For the sake of simplifying notation, from now on we will suppress
the subscript on the one-particle distribution function $f_1$ and use instead simply $f$.

A kinetic equation is a closed equation for the one-particle
distribution function $f$. We see from the BBGKY hierarchy that the first equation for $s=1$ depends on $f_2$
while the equation for $f_2$ depends on $f_3$ and so on. A useful approach consists in defining
the correlations functions $g_s({\bf 1},\ldots,{\bf s};t$) from the following
cluster expansion for the $s$-particle distributions:
\begin{eqnarray}
 f_2({\bf 1},{\bf 2};t) & = & f({\bf 1};t)f({\bf 2};t)+g_2({\bf 1},{\bf 2};t),
\\
 f_3({\bf 1},{\bf 2},{\bf 3};t) & = & f({\bf 1};t)f({\bf 2};t)f({\bf 3};t)+f({\bf 1};t)g_2({\bf 2},{\bf 3};t)
+f({\bf 2};t)g_2({\bf 1},{\bf 3};t)
\nonumber\\
 & & +f({\bf 3};t)g_2({\bf 1},{\bf 2};t)+g_3({\bf 1},{\bf 2},{\bf 3}.t),
\label{clusterexp}
\end{eqnarray}
By replacing this expansion into the BBGKY hierarchy in Eq.~(\ref{bbgky2}) we obtain a hierarchy of equations for
the reduced distribution functions. The two first equations of the BBGKY hierarchy now read
\begin{equation}
\partial_{t}f(\bold{1};t)=\hat L_{1}^{0}f(\bold{1};t)+\int \dd\bold{2}\,\hat L'_{12}\,f(\bold{1};t)\,f(\bold{2};t)+
\int \dd\bold{2}\,\hat L'_{12}\,g_{2}(\bold{1},\bold{2};t),
\label{eq:1}
\end{equation}

\begin{eqnarray}
\frac{\partial}{\partial t}g_2(\bold{1},\bold{2};t) & = & \big[{\textstyle \hat L_{1}^{0}+\hat L_{2}^{0}\big]g_{2}(\bold{1},\bold{2};t)
+\hat L'_{12}\big[g_{2}(\bold{1},\bold{2};t)+f(\bold{1};t)\,f(\bold{2};t)\big]}\nonumber \\
 &  & +\int \dd\bold{3}\left\{\hat L'_{13}f(\bold{1};t)g_{2}(\bold{2},\bold{3};t)+\hat L'_{23}f(\bold{2};t)g_{2}(\bold{1},\bold{3};t)\right.
\nonumber \\
 &  & \left.+(\hat L'_{13}+\hat L'_{23})[f(\bold{3};t)g_{2}(\bold{1},\bold{2};t)+g_{3}(\bold{1},\bold{2},\bold{3},t)]\right\}.
\label{eq:6}
\end{eqnarray}
An example of special interest here of a kinetic equation deduced from the BBGKY hierarchy is the Balescu-Lenard equation. 
It is obtained by neglecting the three-particle correlation function $g_3$ and the direct interaction 
term $\hat L'_{12}g_{2}(\bold{1},\bold{2};t)$ in Eq.~(\ref{eq:6}). The resulting equation for $g_2$  is then 
solved and the result inserted into Eq.~(\ref{eq:1})~\cite{key-10-2}. The Balescu-Lenard is, then, obtained after a markovianization 
of a time convolution integral resulting from the previous step.

The integration domain over $\vec{r}_{i}$
is the volume $V$ of the system, but in view of the thermodynamic
limit, $N\rightarrow\infty$, $V\rightarrow\infty$, $N/V=n={\rm constant}<\infty$,
considered here, the domain is assimilated to $\mathbb{\mathbb{R}^{\mathrm{3}}}$.
The thermodynamic limit should not be confused with the fluid limit
in which $N\rightarrow\infty$, $m\rightarrow0$, $\gamma\rightarrow0,$
$Nm={\rm constant}<\infty$, $N\gamma={\rm constant}<\infty$. The fluid limit
removes the discrete character of the particles. Since we are interested
in a phenomenon related to the discreteness of particles, the thermodynamic
limit is taken here. For a discussion of the two limits see reference~\cite{key-11-1}.

Our concern is limited to weakly coupled systems, i.e. systems for
which $\Gamma\equiv U/K\ll1$, where $U$ and $K$ are the average potential
and kinetic energies per particle, respectively. 
The quantity $U\equiv\left|U(\delta)\right|=\left|\gamma\right|/\delta$
represents the potential energy between two particles at the average
distance $\delta=n^{-1/3}$ between nearest neighbors. The weak coupling
condition is, thus, $\Gamma=\left|\gamma\right|n^{1/3}/K\ll1$. This
condition only imposes that the binary interactions are weak in average
and does not exclude the local existence of strongly interacting
particles. We also assume that the system is dilute. For such weakly coupled systems, in the current theories,
the third term $C\equiv\int \dd\bold{2}\,L'_{12}\,g_{2}(\bold{1},\bold{2};t)$
in Eq.~(\ref{eq:1}) is shown to contain the effects of binary
collisions~\cite{key-10-1}. Under these collisions, the system irreversibly
relaxes towards thermodynamic equilibrium in a time $t_{r}\sim t_{s}\Gamma^{-3/2}$~\cite{key-11},
the Balescu-Lenard collisional relaxation time, where $t_{s}=\sqrt{m/\left|\gamma\right|n}$.
The short time-scale $t_{s}$ represents for plasmas the plasma oscillations
period, and for gravitational systems the collapse time or Jeans instability
time~\cite{key-22}. This time characterizes the collapse process
that starts in a given region when the internal pressure becomes smaller
than the gravitational attraction. Since we assumed $\Gamma\mathit{\ll\mathrm{1}}$,
$t_{r}$ is very large. Hence, for times $t\ll t_{r}$, the effect of the term $C$ can be neglected.
This reduces Eq.~(\ref{eq:1}) to the well-known Vlasov equation.

However, for interaction forces that diverge as $1/r^{2}$ at small
distances, which is the case considered here, the above reasoning is dubious.
Indeed, the domain of the integral over $\vec{r}_{2}$ that appears
in $C$ contains the origin located at the position $\vec{r}_{1}$
of the particle $\mathit{\mathbf{1}}$ where the divergence of the
force occurs. The contribution to that integral in $C$ of a small
volume around the origin, hence, must be carefully studied. To do
so, we divide the integration domain of the integral over $\vec{r_{2}}$
in $C$ in two parts: a small open ball $S_{1}$ of radius $d$ centered
at particle $\mathbf{1}$, and the rest of the space, $\mathbb{R}\mathrm{^{3}\mathrm{\backslash\mathrm{\mathit{S}}_{1}}}$.
The radius $d$ is such that the average interaction energy between
any particle $2$ located in that sphere and particle $\bold{1}$
at the center is larger than the sum of the average kinetic energies
of the two particles $K$. We, thus, have $\left|\gamma\right|/Kd=1$
or $d=\left|\gamma\right|/K$, the Landau length. We also assume that the typical macroscopic inhomogeneity
length $L_{H}$ is much larger than $d$, $d\ll L_{H}$ and also
that $\delta\ll L_{H}$. Combined with $\Gamma\ll1$, this yields
$d\ll\delta\ll L_{H}$.

The splitting of $C$ leads to
\begin{equation}
C=I_{1}+I_{2}\label{eq:2}
\end{equation}
with,
\begin{equation}
I_{1}=\int_{S_{1}}d^{3}r_{2}\int \dd^{3}v_{2}\,\hat L'_{12}\,g_{2}(\vec{r}_{1},\vec{v}_{1},\vec{r}_{2},\vec{v}_{2};t)\label{eq:3}
\end{equation}
and,
\begin{equation}
I_{2}=\int_{\mathbb{R\mathrm{^{3}\mathrm{\backslash\mathrm{\mathit{S}}_{1}}}}}d^{3}r_{2}\int \dd^{3}v_{2}\,\hat L'_{12}\,g_{2}(\vec{r}_{1},\vec{v}_{1},\vec{r}_{2},\vec{v}_{2};t)\label{eq:4}
\end{equation}
Obviously, the norm of the interaction force $F_{12}\equiv||\vec{F_{12}}||$
contained in $\hat L'_{12}$ is large in $I_{1}$ while it is small
in $I_{2}$. As for $I_{\mathbf{\mathrm{2}}}$, by construction it
mainly involves weak binary interactions. This is precisely the condition
required to derive the Balescu-Lenard equation~\cite{key-10-2}.
Using the splitting in Eq.~(\ref{eq:2}), Eq.~(\ref{eq:1}) becomes
\begin{eqnarray}
\partial_{t}f(\bold{1};t) & = & \hat L_{1}^{0}f(\bold{1};t)+\int \dd\bold{2}\, \hat L'_{12}\,f(\bold{1};t)\,f(\bold{2};t)
\nonumber\\
 &  & +\int_{S_{1}}d^{3}r_{2}\int \dd^{3}v_{2}\, \hat L'_{12}\,g_{2}(\vec{r}_{1},\vec{v}_{1},\vec{r}_{2},\vec{v}_{2};t)+I_{2}.
\label{eq:60-1}
\end{eqnarray}
Let us stress that in the standard derivations of the kinetic equations,
the discussion of $I_{1}$ is eluded. It is simply replaced by a cut-off
at short distances in the integral over $r_{2}$ in $C$ in order
to avoid the divergence at the origin. Thus, in these theories $C$
reduces to $I_{2}$ with a phenomenological cut-off. However, in the sequel
we show that $I_{1}$ does not diverge (see Appendix~\ref{appendixa2})
and, consequently, the kinetic equation should contain this term, which we now proceed to derive.

A closed form for the kinetic equation requires the determination of the two-particles correlation 
function $g_2$ in Eq.~(\ref{eq:60-1}) that appears in both $I_{1}$ and $I_{2}$.
However, Eq.~(\ref{eq:6}), obviously, is not closed as it contains
the three-particles correlation $g_{3}$. The time evolution of the latter
is coupled to the equation for $g_{4}$ and so on, generating a
whole hierarchy of coupled equations. Approximations must therefore
be made in order to solve Eq.~(\ref{eq:6}). The
domain of definition of $g_{2}$ in the position space
being different in the two integrals involved in $I_{1}$ and $I_{2}$,
the approximation schemes to solve Eq.~(\ref{eq:6}) are necessarily different
in each case. In order to obtain the correlation $g_{2}$ appearing
in $I_{1}$, Equation~(\ref{eq:6}) must be considered with the
constraint $\left\Vert \vec{r}_{2}-\vec{r}_{1}\right\Vert <d$ while
for the correlation $g_{2}$ contained in $I_{2}$, the constraint
would be $\left\Vert \vec{r}_{2}-\vec{r}_{1}\right\Vert \geq d$.
We will focus on the treatment of Eq.~(\ref{eq:6}) with the
condition $\left\Vert \vec{r}_{2}-\vec{r}_{1}\right\Vert <d$ as its
treatment with the complementary condition is well-known and brings
$I_{2}$ into the form of the Balescu-Lenard collision term with the cut-off 
imposed by that condition. As discussed
in details in Appendix~\ref{appendixa1}, the conditions $\left\Vert \vec{r}_{2}-\vec{r}_{1}\right\Vert <d$,
$t\ll t_{r}$ along with $d\ll\delta\ll L_{H}$ allow for truncating
and greatly simplifying the hierarchy in this case. Indeed, it turns
out  (see Appendix~\ref{appendixa1}) that Eq.~(\ref{eq:6}) reduces to the closed equation:
\begin{equation}
\partial_{t}g_{2}(\bold{1},\bold{2};t)=\hat L'_{12}\big[g_{2}(\bold{1},\bold{2};t)+f(\bold{1};t)\,f(\bold{2};t)\big]\label{eq:2.7}
\end{equation}
As can be observed, this equation corresponds to Eq.~(\ref{eq:6})
where the first and third terms on the right-hand side have been neglected. The reasons of
these approximations are the following. The first term in the right-hand side of Eq.~(\ref{eq:6})
represents the contribution of free motion. In a first approximation, 
in the sphere $S_{1}$, this term is negligible compared to the direct interaction
term (the second term) which is large in the domain $\left\Vert \vec{r}_{2}-\vec{r}_{1}\right\Vert <d$.
Indeed, the trajectories of the two interacting particles in the small sphere
are strongly curved and the contribution of rectilinear inertial motion to them is negligible in a first approximation.
The third term in the right-hand side of Eq.~(\ref{eq:6}) contains the effects of the interactions of
any particle $3$ with the given couple of particles $\mathbf{1}$
and $\mathbf{2}$. This term is shown to be
negligible compared to the second term due to the conditions of global
weak coupling and dilution we assumed for the system.

The solution of this equation is obtained in Appendix~\ref{appendixa2}, and after introducing it in Eq.~(\ref{eq:3}) one gets:
\begin{equation}
I_{1}=-\frac{1}{5}\left(\frac{2\pi\left|\gamma\right|}{m}\right)^{3/2}\int_{0}^{t}\frac{d\tau}{\sqrt{\tau}}\,\,n(\vec{r}_{1};t-\tau)\left(-\triangle_{\vec{v}_{1}}\right)^{3/4}\,\,f(\vec{r}_{1},\vec{v}_{1};t-\tau)\label{eq:8}
\end{equation}
where $n(\vec{r}_{1};t)\equiv\int \dd^{3}{v}f(\vec{r_{1}},\vec{v};t)$
is the local number density. The fractional power $3/4$ of the Laplacian
operator in the velocity variable is defined by 
\begin{equation}
\left(-\triangle_{\vec{v}_{1}}\right)^{3/4}e^{i\vec{\zeta}_{1}\cdot\vec{v}_{1}}\equiv\vec{(\zeta}_{1}.\vec{\zeta}_{1})^{^{3/4}}e^{i\vec{\zeta}_{1}\cdot\vec{v}_{1}}=\zeta_{1}^{3/2}e^{i\vec{\zeta}_{1}\cdot\vec{v}_{1}}\label{eq:14-2}
\end{equation}
and by using the Fourier integral representation of $f(\vec{r}_{1},\vec{v}_{1};t-\tau)$
with respect to the velocity.

In deriving expression~(\ref{eq:8}) we supposed a vanishing binary
correlation $g_{2}$ at time zero. The supplementary term that would
appear for non-vanishing initial correlations is given and discussed
in the Appendix~\ref{appendixa2}. We explain in the same appendix why that supplementary
term does not change the main conclusions of this work. For this reason,
the present article focuses on the structure of the main result, that
is, on expression~(\ref{eq:8}).

Finally, Eqs.~(\ref{eq:8}) and~(\ref{eq:60-1}) allow to write a
closed equation for the one-particle distribution function $f$, i.~e.\ the final form
of our kinetic equation:
\begin{eqnarray}
\partial_{t}f(\vec{r_{1},}\vec{v_{1}};t) & = &-\vec{v}_{1}\cdot\vec{\nabla}_{1}f(\vec{r_{1}},\vec{v_{1};}t)
-\frac{1}{m}\int \dd^{3}{r_{2}}\int \dd^{3}{v_{2}}\,\vec{F}_{12}\cdot(\frac{\partial}{\partial\vec{v}_{1}}
-\frac{\partial}{\partial\vec{v}_{2}})\,f(\vec{r_{1}},\vec{v_{1}};t)\,f(\vec{r_{2}},\vec{v_{2}};t)\mbox{\ensuremath{\mbox{}}}\nonumber\\
 & & -\frac{1}{5}\left(\frac{2\pi\left|\gamma\right|}{m}\right)^{3/2}\int_{0}^{t}\frac{d\tau}{\sqrt{\tau}}\,\,n(\vec{r}_{1};t-\tau)\left(-\triangle_{\vec{v}_{1}}\right)^{3/4}\,\,f(\vec{r}_{1},\vec{v}_{1};t-\tau)+I_{2}.
\label{eq:15-2}
\end{eqnarray}
Equation~(\ref{eq:15-2}) is the usual kinetic equation modified
by the addition of a new term, the third term in the right-hand-side.
This new contribution is nonlinear in $f_1$ due to the presence
in it of $n(\vec{r}_{1};t-\tau)$ and involves a fractional power
$\frac{3}{4}$ of the velocity Laplacian. The time integral is a fractionary
iterated integral of order $\frac{1}{2}$ as explained in the next
section. As mentioned above, $I_{2}$ represents the Balescu-Lenard collision
term with a small distance cut-off defined by the Landau length $d$.
This contribution is also derived from the BBGKY hierarchy. However,
the truncation scheme of that hierarchy is now different from the
previous derivation leading to the term~(\ref{eq:8}) since the possible
distances between particles $\mathbf{1}$ and $\mathbf{2}$ are larger
than $d$. The effects of interactions may, thus, be considered as
weak as compared to the free motion. As a consequence, the terms in the second equation~(\ref{eq:6})
of the BBGKY have not the same relative importance as in the derivation
leading to~(\ref{eq:8}).

It is important to note that Eq.~(\ref{eq:15-2}) involves an integral on time with a time delay in the integrand and, consequently,
the above kinetic equation is not Markovian. In contrast with the Markovianization that naturally occurs in the Boltzmann  or the
Balescu-Lenard collision terms, here the slow decay of the $\frac{1}{\sqrt{\tau}}$ kernel forbids it. Indeed, due to this slow decay,
the time-delay $\tau$ in the arguments of the local density and of the one-particle distribution appearing in that integral can not be neglected.
Moreover, in contrast with the two previously mentioned collision terms, this slow decay also forbids the extension to infinity of the upper boundary of the time integral, thereby, leaving an intrinsic time dependency in the new term.\\

However, the precise form of that non-Markovianity has a peculiar and deep mathematical meaning: it causes
the apparition of a fractional order of the time derivative present in the kinetic equation as shown 
in the next chapter (see Eq.~\ref{eq:53}) for the case of uniform systems. We now show that, in the case of spatially uniform systems and for
short times, the general solution to the kinetic equation is a velocity distribution with a long tail in $1/v^{5/2}$.\\

Before closing this chapter, let us mention two remarks. First, one of us~\cite{yassinthesis}
calculated the first corrections to the total neglect of the free motion term in Eq.~(\ref{eq:6}) in the equation for $g_{2}$. 
To that purpose, he developed a perturbation expansion in this small term. The first order 
contributions was found to vanish exactly.  The next correction involved terms with integer powers of
the velocity Laplacian. Second, we reproduced the same reasoning as followed in this chapter but starting from the quantum BBGKY
equation for the reduced Wigner functions. This led to a very similar contribution to the kinetic equation. Moreover, this contribution tends to the above new term in the classical limit. These results will be published in a separate article elsewhere.\bigskip{}

\section{Homogeneous systems}
\label{sc3}

Let us consider the particular situation of an initially uniform system. 
Such a configuration can be maintained in globally neutral
plasmas. But for self-gravitating systems it is not a stable situation.
Nevertheless, we limit here our scope to times that are very short
and certainly shorter than the Jeans instability time $t_{s}$. Indeed, the
characteristic time  $t_{f}$ of the new term in the kinetic equation
is of the order of the time a particle moving at the average velocity
$\sqrt{K/m}$ stays in the small sphere of radius $d$. This
implies $t_{f}/t_s=\Gamma^{3/2}\ll1$ which reflects the
condition of average weak coupling.

As to the free motion and Vlasov terms in Eq.~(\ref{eq:15-2}), they exactly vanish for uniform systems.
Moreover, for times much shorter than the collisional relaxation time $t_{r}$
one can also neglect the collision term $I_{2}$. Indeed, one can show that
$t_{f}/t_{r}=\Gamma^{3}\ll1$. This approximation leaves only
the new term in that equation and singles out its effect on the evolution
of the one-particle distribution function.

For uniform systems we have $f(\vec{r},\vec{v,}t)=n\varphi(\vec{v,}t)$,
where $\varphi(\vec{v},t)$ is the velocity distribution at time $t$.
Equation~(\ref{eq:15-2}) then becomes:
\begin{equation}
\partial_{t}\varphi(\vec{v};t)=\,\,-\frac{n}{5}\left(\frac{2\pi\left|\gamma\right|}{m}\right)^{3/2}\int_{0}^{t}\frac{d\tau}{\sqrt{\tau}}\,\,\left(-\triangle_{v}\right)^{3/4}\,\,\varphi(\vec{v};t-\tau).
\label{eq:22}
\end{equation}
Obviously, the uniformity condition made the kinetic equation exactly
linear. One can, thus, find the exact solution to
this equation. This is far from being the case for the general nonlinear
equation in Eq.~(\ref{eq:15-2}). The solution of Eq.~(\ref{eq:22})
is obtained in Appendix~\ref{appendixb} and amounts to a convolution integral
in the velocity between the initial velocity distribution and the inverse Fourier
transform of the Mittag-Leffler function $E_{3/2}(z)$ of index $3/2$~\cite{key-13}:
\begin{equation}
\varphi(\vec{v};t)=\int\frac{d^{3}\zeta}{(2\pi)^{3}}e^{i\vec{\zeta}\cdot\vec{v}}\,\,
\tilde{\varphi}(\vec{\zeta};0)\,\,E_{3/2}\left(-\frac{n\pi^2}{5}\left[\frac{2\left|\gamma\right|}{m}\right]^{3/2}\,\,\zeta^{3/2}\,\,t^{3/2}\right).
\label{eq:52}
\end{equation}
For all initial distributions with finite second moments, the above
convolution gives a long-tailed distribution with tail $1/{v}^{5/2}$~\cite{key-7-1-1}
where $v$ is any component of the velocity vector $\vec{v}$.
This result has some experimental confirmations as discussed below
and, as shown in the next section, it is confirmed by molecular
dynamical simulations.

Let us now put the long tail in the velocity distribution obtained
in this section in a more general context. Non-Gaussian and, more
particularly, long-tailed velocity distributions are observed and/or
predicted in various far-from-equilibrium macroscopic systems. For
systems with short-range interactions, high-velocity tails have been
predicted for a two-dimensional gas of Maxwell molecules under uniform
shear flow~\cite{key-1-1} and for non-equilibrium steady-states
of inelastic gases~\cite{key-1-2}.

As shown in Appendix~\ref{appendixb}, for large values of $v$ the inverse Fourier transform
of the function $E_{3/2}$ tends toward the tail of a Lévy distribution of index
$3/2$~\cite{key-13,key-2}. The latter has an algebraic tail in $1/v^{5/2}$.
It turns out that in experiments with focused ion
beams~\cite{key-1}, the observed transverse velocity distribution
of the ions is a symmetric Lévy-stable distribution of stability index
$3/2$~\cite{key-2} with a long tail in $1/v^{5/2}$ . In another
domain, long tails in the velocity distribution are  invoked as a possible
explanation of some properties in the process of ionization in gases
and in nuclear fusion in plasmas. The rates of these processes are
very sensitive to the number of high velocity particles in the system.
In these experiments the measured fusion or ionization rates are often significantly larger
than those predicted using a Gaussian velocity distribution~\cite{key-3}.
These discrepancies led Ebeling, Romanovsky and Sokolov
to replace the Gaussian velocity distribution by a convolution between
a Gaussian and a symmetric Lévy of index $3/2$~\cite{key-3,key-4}.
The rates calculated with this new distribution came closer to the
measured ones~\cite{key-3}. These authors based their reasoning on
the theory of Holtsmark~\cite{key-5} who first demonstrated for
plasmas that under certain conditions the distribution of the total
field exerted on any particle by all the other particles in the system
is a Lévy distribution of index $3/2$.

Though still controverted, similar distributions for the peculiar
velocities of the galaxies are proposed for large-scale systems of
galaxies. Some authors suggest that the distribution of the measured peculiar
velocity has a long tail in $1/v^{2.1}$~\cite{key-7}. The difference
between the exponent $2.1$ of the observed velocity tail and the
exponent $2.5$ of the tail of a true Lévy-$3/2$ distribution could
be related to the existence of power-law correlations in the spatial
distribution of matter~\cite{key-8,key-9,key-11-1}.
One should also stress the fact that Chandrasekhar and von Neumann~\cite{key-6}
derived a distribution for the total gravitational field in a homogeneous
self-gravitating system that is identical to the Holtsmark distribution
for plasmas. Another common aspect of both the Holtsmark and the Chandrasekhar
distributions is that they are obtained as generalized Central-Limit
theorems~\cite{key-2} in the thermodynamic limit in which the total
number of particles tends to infinity.

Before closing this chapter, a last result is worth to be reported.
Equation~(\ref{eq:22}) can be cast into a particularly elegant form.
The fractional iterated integral operator of order $\alpha$ acting
on a function $F(t)$ is defined by~\cite{key-13}:
$$
J_{t}^{\alpha}f(t)\equiv\frac{1}{\Gamma(\alpha)}\int_{0}^{t}\tau^{\alpha-1}F(t-\tau).
$$
Up to a factor $1/\sqrt{\pi}$ and for the particular value $\alpha=1/2$,
this is just the integral operator on variable $\tau$ appearing in
the right hand side of Eq.~(\ref{eq:22}). The Riemann-Liouville
fractional derivative of order $1/2$ in the time variable $t$ is
defined as~\cite{key-13}: $D_{t}^{1/2}\equiv\frac{d}{dt}\circ J_{t}^{1/2}$.
Now, let us apply $D_{t}^{1/2}$ on both sides of Eq.~(\ref{eq:22}).
Using the following group properties, $J_{t}^{1/2}\circ J_{t}^{1/2}=J_{t}^{1}$,
$\frac{d}{dt}\circ J_{t}^{1}=I$, $\frac{d}{dt}\circ D_{t}^{1/2}\equiv D_{t}^{3/2}$,
where $I$ is the identity operator, Equation~(\ref{eq:22})
transforms into 
\begin{equation}
D_{t}^{3/2}\varphi(\vec{v};t)=-\frac{n\pi^2}{5}\left(\frac{2|\gamma|}{m}\right)^{3/2}(-\triangle_{\vec{v}_{1}})^{3/4}\,\,\varphi(\vec{v};t).
\label{eq:53}
\end{equation}
To our knowledge, this is the first time such a fractional kinetic
equation is derived from the basic principles of Statistical Mechanics.\bigskip{}

\section{Numerical simulations}
\label{sc4}

The simulations we present here were confronted with two main difficulties.
The first was the extreme difficulty to perform 3D molecular dynamical
simulations of a sufficiently large number of particles interacting
through a potential in $1/r$ or via a regularized potential as
close as possible to the former. Such simulations are extremely time-consuming. 
The second difficulty was to find a theoretical prediction
that could be recognizable and measurable on the simulations. Since
the general kinetic equation~(\ref{eq:15-2}), due to its nonlinearity, has
not been solved up to now, we had to simulate physical
conditions that are the nearest possible to the unique situation
for which the exact velocity distribution is known, i.~e.\
a spatially uniform system. The exact solution of the kinetic equation in that
case was obtained in the previous chapter.

In order to be in conformity with neglecting the
collisional term $I_{2}$, we had to limit the simulations to times
much smaller than the collisional relaxation time $t_{r}$. Moreover,
in order to be coherent with discarding the free motion and the
Vlasov terms the simulation had to start from an
initial uniform initial condition, and remain as near as possible from
spatial uniformity. For a self-gravitating system, this
limited us to simulation times that are much smaller than the Jeans
time $t_{s}$. As shown in the previous section, for
such simulation conditions, the predicted velocity distribution is
characterized by a long-tail in $1/v^{5/2}$. Hence, the main marker
we needed to measure on these simulations was the exponent of the
tail. We also chose initial conditions with no spatial and velocity
correlations as Eq.~(\ref{eq:22}) was obtained assuming this condition.
\smallskip{}

\begin{figure}[h]
\begin{center}
\includegraphics[scale=0.3]{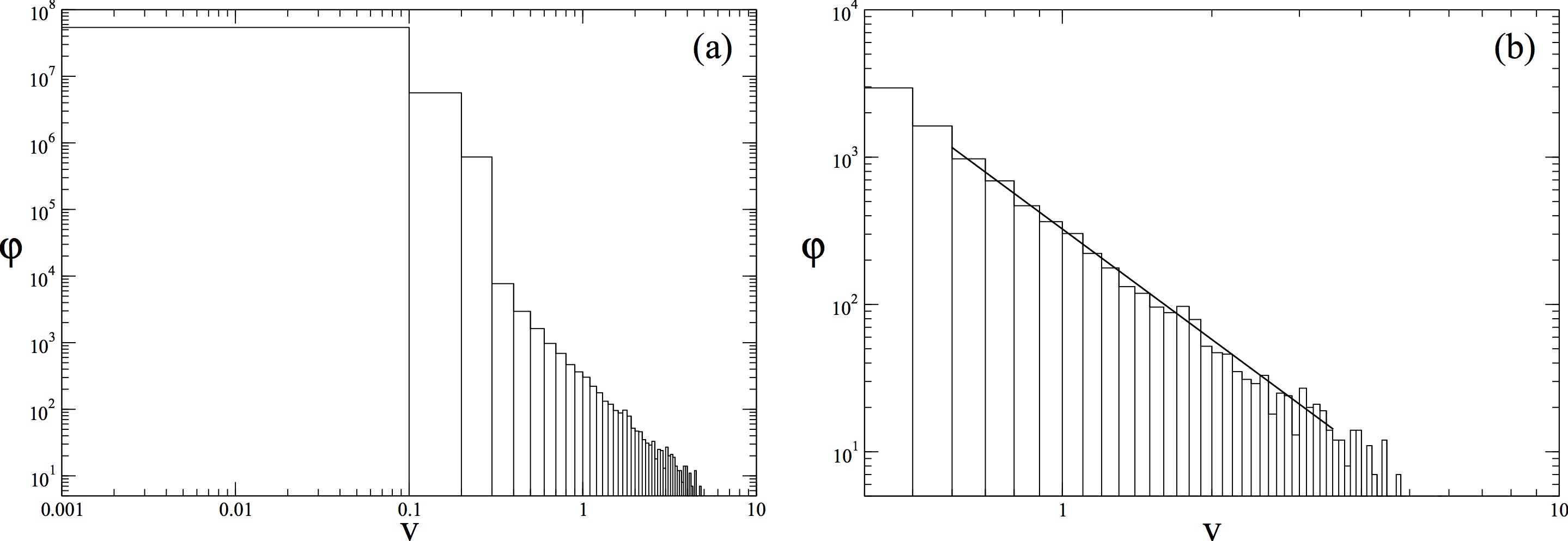}
\end{center}
\caption{(a) Log-log histogram of distribution of any component $v$
of velocity $\vec{v}$ for regularization parameter $\varepsilon=10^{-16}$,
total number of particles $N=131,072$, number of realizations $=100$
and all particles initially at rest. The initial spatial distribution is uniform
in a sphere of radius $R=1.28$. Time step $=2\times10^{-7}T$, total
run time $=3\times10^{-6}T$ with $T=(nGm)^{-1/2}$, $G$ the gravitational
constant.
(b) Zoom on the tail of the velocity distribution given in (a). The
thick straight line is the result of a linear regression on the tail
with slope $\alpha=-2.49$, standard error $=0.13$ and correlation coefficient
$-0.97$.}
\label{fig1}
\end{figure}

More precisely, we simulated a 3D gravitational system of 131,072
identical point-like classical particles using a GPU implementation of a fourth order symplectic
integrator~\cite{yoshida,key-12}.
The initial distribution of the particles was spatially uniform in
a spherical volume with all the particles at rest and without space
boundaries: The system was open and particles could escape. The interaction
potential $\gamma/r$ was regularized as $\gamma/\sqrt{r^{2}+\varepsilon^{2}}$
in order to avoid divergences in the numerical integration. The statistical
significance was increased by performing 100 realizations (runs) and
collecting the final values of the velocities of all particles.
After a very short time, long tails proportional
to $1/v^{\alpha}$ developed in the distribution for any
component $v$ of the velocity vector $\vec{v}$ as can be
seen in Figures~\ref{fig1}a and~\ref{fig1}b.

\begin{figure}[h]
\includegraphics[scale=0.4]{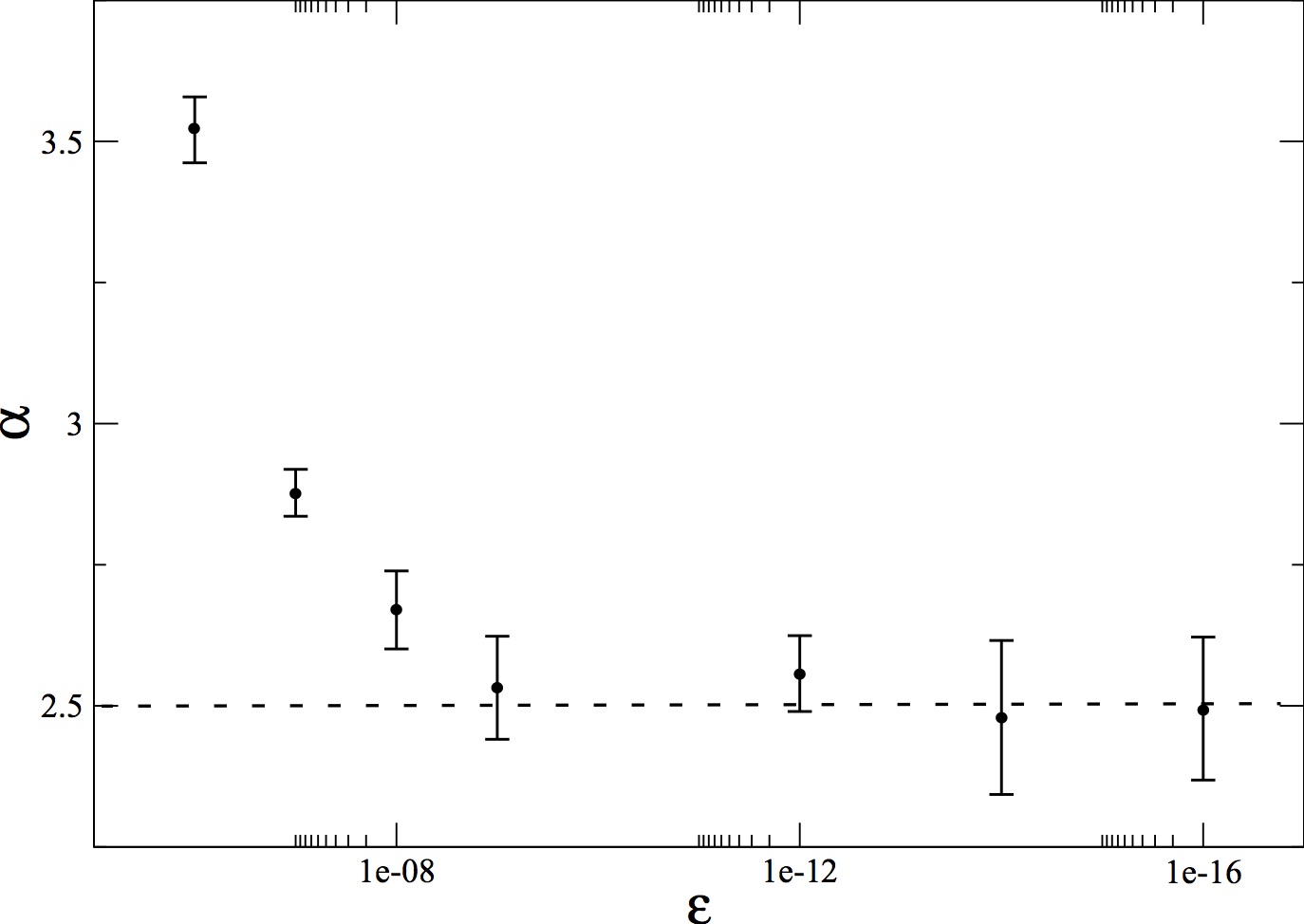}

\caption{Semi-log graph of the tail exponent $\alpha$ in $1/v^{\alpha}$
as a function of the regularization parameter $\varepsilon$. Each
point is obtained from a linear regression on velocity data obtained
from 100 realizations (runs) as in Figure 1.b. Error bars correspond
to the standard deviation obtained from the linear regression and
the dashed line is the theoretical value at $\alpha=2.5$.}
\label{fig2}
\end{figure}

Figure~\ref{fig2} shows the exponent $\alpha$ for decreasing values
of the regularization parameter $\varepsilon$ and the same
initial conditions as in Fig.~\ref{fig1}. Each point corresponds to the average over 100
runs (realizations). As observed, $\alpha$ smoothly approaches the
theoretical value $5/2$ as $\varepsilon$ gets smaller. 

The tail of the distribution is very sensitive to the behavior of
the interaction force at small distances and, consequently, to $\varepsilon$.
The expected behavior is thus observed only for very small values
of $\varepsilon$. Note that the errors (from $5\%$ to $10\%$) on
$\alpha$ should not appear as a great concern as they could be greatly
reduced by increasing either the number of particles in the simulation
or the number of realizations used for statistical significance. Indeed,
the tail of a distribution is much more sensitive to the finiteness
of $N$ than the bulk of the distribution, and numerical simulations (as well as experiments) 
are bound to involve a finite number of particles. As a consequence,
the number of particles with very large velocities, i.~e.\ those that
constitute the end of the tail of the distribution,
represents only a small fraction of $N$. The shape of the tail, thence,
is subject to large fluctuations at its end. Nevertheless, the long
tail property essentially persists. This persistence is explicitly
verified in Figure 2 where it is seen that for decreasing values of
the regularization parameter, the tail exponent of the velocity distribution
rapidly decreases and stabilizes around the predicted value $5/2$.\\

A last remark before ending this chapter. The choice of very peculiar initial conditions
in our simulations by no means implies the inexistence of the new term in time evolutions
with other initial conditions. However, in many of them, it would be difficult to recognize its effects
in the simulation results as long as one does not have theoretical predictions for such initial
conditions.\bigskip{}

\section{Long tails and conserved quantities}
\label{sc5}

At this level one should discuss the physical soundness of the long-tailed
distribution we found in the case of uniform systems. Indeed, its
second moment and all of its higher order moments diverge. Therefore,
in principle, the average kinetic energy obtained from it diverges.
This divergence represents the fact that particles that are very close
from each other acquire very large accelerations and velocities under
the action of the divergent interaction force. In reality, as can be seen on 
equation~(\ref{eq:18-1}), the function $J$ that appears in equation~(\ref{eq:15-1}) 
can be expanded in a convergent series. That series includes a first term that is a 
fractional power of  $\zeta$ followed by a series of terms proportional to even integer powers of
$\zeta$. In the inverse Fourier transform the first term produces the fractional power 
of the velocity Laplacian and the remaining terms give integer powers of the same Laplacian.
Hence, the new fractional term~(\ref{eq:8}) in the kinetic equation derives from the first term in that series.
But in writing it, we neglected the remaining terms that play a growing role for longer times. These terms being
proportional to integer powers of the velocity Laplacian will produce diffusive behavior in the velocity space. These
diffusive effects, in turn, will progressively transform the asymptotic behavior of the algebraic velocity tail into a Gaussian tail. 
That Gaussian tail will appear for increasingly smaller values of the velocity as time goes on at the expense of the algebraic tail.\\
Therefore, if one does not take into account explicitly the corrections in integer powers of the velocity Laplacian
in the kinetic equation, though nothing forbids to do it, one must at least truncate the long tail in the velocity distribution that solves it. 
Consequently, the second velocity moment exists and the question
of the convergence of the kinetic energy is solved\\
As for the conservation of matter and momentum, they correspond to the zero and first orders moments.
These moments exist for the long-tailed distribution even without truncation and we can derive
them directly from Eq.~(\ref{eq:15-2}). The derivation of their conservation by that kinetic equation as well as that of the total energy
raises no difficulty .\bigskip{}

\section{Discussion and perspectives}
\label{sc6}

The new term in the kinetic equation~(\ref{eq:15-2}) is of order
$1/N$ compared to the Vlasov term. However, in uniform or near-uniform
spatial configurations, while the Vlasov term vanishes or is very
small, this term remains finite and, consequently, cannot be neglected.
In fully inhomogeneous states, furthermore, though small compared
to the Vlasov term, this new term might have some consequences. Indeed,
different finite samples of $N$ point particles obeying a statistical
distribution with an algebraic tail usually have very different standard
deviations (diverging with $N$). As a consequence, this new term
may have a measurable influence on the quasi-stationary state that
appears after the violent relaxation process~\cite{key-14,key-15}.
Also, its magnitude in $N$ is of the same order as the collisional
term $I_{2}$. The latter becomes important after long times of the
order of $t_{r}$. Therefore, an extension of the present approach
to the relaxation time-scale might be of interest. Much longer simulation
runs are required to study how the long tails disappear and how the
new term affects the evolution of the system on longer time scales.

On the other hand, it would be worth investigating the effects of the divergence of the potential
at short distances in one-dimensional models, e.~g.\ the self-gravitating ring model~\cite{sota}.
In this model the Balescu-Lenard collisional integral vanishes for homogeneous states~\cite{sano,scaling}
and the contribution of the fractional term (if it exists for this model) should be more easily observed
on the simulation results.

In our derivation of the kinetic equation~(\ref{eq:15-2}) we assumed
a vanishing initial binary correlation function (see Appendix~\ref{appendixa2}).
The effect of a non-vanishing initial correlation is derived
in Appendix~\ref{appendixa2} and
should be added to the kinetic equations~(\ref{eq:15-2}) and~(\ref{eq:22}).
It introduces a source term in the kinetic equation and its solution
involves a time convolution between this source term and
the propagator of Eq.~(\ref{eq:15-2}) or of Eq.~(\ref{eq:22})
in the uniform case. Limiting our ambition to the latter and using
a theorem in reference~\cite{key-7-1-1}, one shows for uniform systems
and for short times that the tail of the resulting distribution still is
proportional to $1/v^{5/2}$ for a large class of initial
correlations. Furthermore, the contribution of the correlations in the initial
conditions to the kinetic equation is expected to rapidly decay with time
(see discussion on this point in Ref.~\cite{key-10-2}).

The present theory extends without difficulty to systems with more
general interaction forces that behave as $1/r^{2}$ mainly at short
distances as, for instance, systems interacting via a Yukawa-type
or Debye-screened potential $\gamma e^{-r/\lambda}/r$ . Moreover,
if in addition one has $\lambda\ll L_{H}$, i.e. the interaction is
short-ranged, then the Vlasov term is negligible~\cite{key-10-2}
and the supplementary term we derived is dominant in the kinetic equation.
Yukawa-like effective potentials also play an important role in nuclear
physics and, more particularly, in heavy-ion collisions such as those
occurring in the large accelerators. However, the nuclear effective
interaction is more complex than the Yukawa potential~\cite{key-16}.
The former depends on the spins and isospins of the two interacting
particles. In some spin and isospin states the potential diverges
as $1/r$ at short distances, in others it behaves as $1/r^{3}$.
Quantum effects are, thus, important in heavy-ion interactions and
would require the quantum version of our approach or, at least, its semi-classical 
expansion.

Another direction worth to be investigated concerns the effects of
the dimensionality of the space and of the exponent c of the interaction
potential $1/r^{c}$ on the various terms of the kinetic equation
and on their convergence. Such a study can be found in Ref.~\cite{key-10-1}
but without taking into account the new term we derived here.

\begin{acknowledgments}
The authors are greatly indebted to Drs.\ J.~Wallenborn (ULB) and A.~Figueiredo
(UnB) for the many deep and fruitful discussions during the elaboration
of the present work.
\end{acknowledgments}

\appendix

\section{Derivation of the kinetic equation~(\ref{eq:15-2}) }
\label{appendixa}

\subsection{Truncation of the BBGKY hierarchy for $\left\Vert \vec{r}_{2}-\vec{r}_{1}\right\Vert <d$}
\label{appendixa1}

Let us analyze the integral term in Eq.~(\ref{eq:6}). We denote it by $K$ 
\begin{eqnarray}
K & =\int_{\mathbb{R}^{3}}\dd^{3}{r_{3}}\int_{\mathbb{R}^{3}}\dd^{3}{v_{3}}\Big\{ \hat L'_{13}f(\bold{1};t)g_{2}(\bold{2},\bold{3};t)
+\hat L'_{23}f(\bold{2};t)g_{2}(\bold{1},\bold{3};t)+\nonumber \\
 & (\hat L'_{13}+\hat L'_{23})\big[f(\bold{3};t)g_{2}(\bold{1},\bold{2};t)+g_{3}(\bold{1},\bold{2},\bold{3};t)\big]\Big\},
\nonumber 
\end{eqnarray}
and is the sum of four contributions $K=K_{1}+K_{2}+K_{3}+K_{4}$
that are defined below. The first one is 
\begin{equation}
K_{1}=\int_{\mathbb{R}^{3}}\dd^{3}{r_{3}}\int_{\mathbb{R}^{3}}\dd^{3}{v_{3}}\hat L'_{13}\,f(\bold{1};t)g_{2}(\bold{2},\bold{3};t)\label{eq:55}
\end{equation}
or more explicitly and with a permutation of integrals 
\begin{equation}
K_{1}=-\frac{1}{m}\int_{\mathbb{R}^{3}}\dd^{3}{v_{3}}\int_{\mathbb{R}^{3}}\dd^{3}{r_{3}}\vec{F}(\vec{r_{1}}-\vec{r}_{3}).(\frac{\partial}{\partial\vec{v}_{1}}-\frac{\partial}{\partial\vec{v}_{3}})f(\vec{r}_{1},\vec{v}_{1};t)g_{2}(\vec{r}_{2},\vec{v}_{2},\vec{r}_{3},\vec{v}_{3};t)\label{eq:56}
\end{equation}
where $\vec{F}(\vec{r})\equiv\gamma\:\vec{r}/r^{3}$ . The part
of the volume integral over $\vec{v}_{3}$ that contains the derivative $\partial/\partial\vec{v}_{3}$
transforms into a surface integral on the surface at infinity in the
sub-space of velocity $\vec{v}_{3}$ and vanishes due to the fact
that $g_{2}(\vec{r}_{2},\vec{v}_{2},\vec{r}_{3},\vec{v}_{3};t)\rightarrow0$
for $v_{3}\rightarrow\infty$~\cite{key-10-2}. Let us make successively
two changes of variable in the integral over $\vec{r}_{3}$: First,
$\vec{r}_{3}\rightarrow\vec{r}=\vec{(r_{1}}-\vec{r}_{3})$ and, second,
$\vec{r}\rightarrow\vec{F}=\gamma\:\vec{r}/r^{3}$. In the
last transformation the volume differential element becomes $d^{3}r=\frac{1}{2}\left|\gamma\right|^{3/2}F^{-9/2}d^{3}F$.
Hence, $K_{1}$ reads now 
\begin{equation}
K_{1}=-\frac{\left|\gamma\right|^{3/2}}{2m}\int_{\mathbb{R}^{3}}\dd^{3}{v_{3}}\int_{\mathbb{R}^{3}}\dd^{3}{F}\,F^{-9/2}\,\vec{F}\thinspace g_{2}(\vec{r}_{2},\vec{v}_{2},\vec{r}_{1}-\left|\gamma\right|^{1/2}F^{-3/2}\vec{F},\vec{v}_{3};t).\frac{\partial}{\partial\vec{v_{1}}}f(\vec{r}_{1},\vec{v}_{1};t)\label{eq:57}
\end{equation}

We now express the integral over $\vec{F}$ in spherical coordinates
$F$, $\theta$, $\varphi$ 
\begin{multline}
\begin{aligned}\end{aligned}
K_{1}=-\frac{\left|\gamma\right|^{3/2}}{2m}\int_{\mathbb{R}^{3}}\dd^{3}{v_{3}}\int_{0}^{\pi}\dd\theta\thinspace sin\theta\int_{0}^{2\pi}\dd\varphi\thinspace\vec{n}(\theta,\varphi)\\
\int_{0}^{\infty}\dd F\thinspace F^{-3/2}g_{2}(\vec{r}_{2},\vec{v}_{2},\vec{r}_{1}-\left|\gamma\right|^{1/2}F^{-1/2}\vec{n}(\theta,\varphi),\vec{v}_{3};t).\frac{\partial}{\partial\vec{v_{1}}}f(\vec{r}_{1},\vec{v}_{1};t)\label{eq:58}
\end{multline}

where $\vec{n}(\theta,\varphi)$ is the unit vector 
\[
\vec{n}(\theta,\varphi)=\left(\begin{array}{c}
sin\theta\thinspace cos\varphi\\
sin\theta\thinspace sin\varphi\\
cos\theta
\end{array}\right)
\]

Finally, the change of variable $F\rightarrow u=F^{-1/2}$ yields
\begin{align}
\begin{aligned}\end{aligned}
K_{1} & =-\frac{\left|\gamma\right|^{3/2}}{m}\int_{\mathbb{R}^{3}}\dd^{3}{v_{3}}\int_{0}^{\pi}\dd\theta
\thinspace sin\theta\int_{0}^{2\pi}\dd\varphi\thinspace\vec{n}(\theta,\varphi)\label{eq:59}\\
 & \int_{0}^{\infty}\dd u\thinspace g_{2}(\vec{r}_{2},\vec{v}_{2},\vec{r}_{1}-\left|\gamma\right|^{1/2}u
\thinspace\vec{n}(\theta,\varphi),\vec{v}_{3};t).\frac{\partial}{\partial\vec{v_{1}}}f(\vec{r}_{1},\vec{v}_{1};t).\nonumber 
\end{align}

Notice that the divergence of the integral over $\vec{r_{3}}$ in
Eq.~(\ref{eq:56}) that could have been expected from the divergence
of the force when $\vec{r}_{3}\rightarrow\vec{r}_{1}$ does not occur
here. Indeed, in Eq.~(\ref{eq:59}) the integral
over $u$ contains only $g_{2}$ which, in turn, is an integrable
function of all its arguments. The last claim comes from the fact
that the two-particles phase-space distribution $f_{2}(\vec{r}_{2},\vec{v}_{2},\vec{r}_{3},\vec{v}_{3};t)$
must be integrable in order to be normalizable. Hence, for all values
of $\vec{r_{1}}$ and $\vec{r_{2}}$ the term $K_{1}$ is finite.
This contrasts with the term $\mathcal{L\equiv}\hat L'_{12}\big[g_{2}(\bold{1},\bold{2};t)+f(\bold{1};t)\,f(\bold{2};t)\big]$
of equation~(\ref{eq:6}) where in $\mathcal{\mathit{\hat L'_{12}}}$
the force diverges for $\vec{r}_{2}\rightarrow\vec{r}_{1}$. Since
Eq.~(\ref{eq:6}) is considered here with the constraint $\left\Vert \vec{r}_{2}-\vec{r}_{1}\right\Vert <d$,
$\mathcal{L}$ is dominant over $K_{1}$. The same argument applies
to $K_{2}$ with the permutation $1\longleftrightarrow2$:
\begin{equation}
K_{2}=\int_{\mathbb{R}^{3}}\dd^{3}{r_{3}}\int_{\mathbb{R}^{3}}\dd^{3}{v_{3}}\hat L_{23}\,f(\bold{2};t)g_{2}(\bold{1},\bold{3};t).
\end{equation}

Using the same changes of variables as above, the term $K_{3}$ 
\begin{equation}
K_{3}=\int_{\mathbb{R}^{3}}\dd^{3}{r_{3}}\int_{\mathbb{R}^{3}}\dd^{3}{v_{3}}(\hat L'_{13}+\hat L'_{23})g_{3}(\bold{1},\bold{2},\bold{3};t),
\label{eq:60}
\end{equation}
becomes 
\begin{align}
K_{3} & =-\frac{\left|\gamma\right|^{3/2}}{m}\frac{\partial}{\partial\vec{v}_{1}}.\int_{\mathbb{R}^{3}}\dd^{3}{v_{3}}\int_{0}^{\pi}\dd\theta\thinspace sin\theta\int_{0}^{2\pi}\dd\varphi\thinspace\vec{n}(\theta,\varphi)\int_{0}^{\infty}\dd u\thinspace g_{3}(\vec{r}_{1},\vec{v}_{1,}\vec{r}_{2},\vec{v}_{2},\vec{r}_{1}-\left|\gamma\right|^{1/2}u\thinspace\vec{n}(\theta,\varphi),\vec{v}_{3};t)\nonumber \\
 & -\frac{\left|\gamma\right|^{3/2}}{m}\frac{\partial}{\partial\vec{v}_{2}}.\int_{\mathbb{R}^{3}}\dd^{3}{v_{3}}\int_{0}^{\pi}\dd\theta\thinspace sin\theta\int_{0}^{2\pi}\dd\varphi\thinspace\vec{n}(\theta,\varphi)\int_{0}^{\infty}\dd u\thinspace g_{3}(\vec{r}_{1},\vec{v}_{1,}\vec{r}_{2},\vec{v}_{2},\vec{r}_{2}-\left|\gamma\right|^{1/2}u\thinspace\vec{n}(\theta,\varphi),\vec{v}_{3};t),
\end{align}
and with a similar argument as for $K_{1}$ and $K_{2}$ one can neglect
$K_{3}$ with respect to $\mathcal{L}$. Finally, let us consider
the term $K_{4}$ 
\begin{equation}
K_{4}=\int_{\mathbb{R}^{3}}\dd^{3}{r_{3}}\int_{\mathbb{R}^{3}}\dd^{3}{v_{3}}(\hat L'_{13}+\hat L'_{23})f(\bold{3};t)g_{2}(\bold{1},\bold{2};t),
\label{eq:62}
\end{equation}
or more explicitly:
\begin{equation}
K_{4}=-\frac{1}{m}\{\int_{\mathbb{R}^{3}}\dd^{3}{r_{3}}\vec{F}(\vec{r_{1}}-\vec{r}_{3})n(\vec{r}_{3};t).\frac{\partial}{\partial\vec{v}_{1}}+\int_{\mathbb{R}^{3}}\dd^{3}{r_{3}}\vec{F}(\vec{r_{2}}-\vec{r}_{3})n(\vec{r}_{3};t).\frac{\partial}{\partial\vec{v}_{2}}\}g_{2}(\bold{1},\bold{2};t).
\label{eq:63}
\end{equation}
The integral $\mathcal{\vec{F}}(\vec{r}_{1})\equiv\int_{\mathbb{R}^{3}}\dd^{3}{r_{3}}\vec{\,F}(\vec{r_{1}}-\vec{r}_{3})\,n(\vec{r}_{3};t)$,
the Vlasov mean force field in Eq.~(\ref{eq:63}), represents
$N$ times the mean force that another particle 3 exerts on particle
1 averaged on the position probability density $p(\vec{r_{3}};t)\equiv n(\vec{r_{3}};t)/N$.
The second integral has the same meaning but with particle 1 replaced
by particle 2. Let us, then, compare $K_{4}$ to the term $\mathcal{L}$
written more explicitly as 
\begin{equation}
\mathcal{L}=-\frac{1}{m}\vec{F}(\vec{r_{1}}-\vec{r}_{2}).(\frac{\partial}{\partial\vec{v}_{1}}
-\frac{\partial}{\partial\vec{v}_{2}})[g(\bold{1},\bold{2};t)+f(\bold{1};t)\,f(\bold{2};t)].
\label{eq:64}
\end{equation}

We must compare the orders of magnitude of $||\vec{F}(\vec{r_{1}}-\vec{r}_{2})||$
and $||\mathcal{\vec{F}}(\vec{r}_{i})||$, $i=1,2$,
for $\left\Vert \vec{r}_{2}-\vec{r}_{1}\right\Vert <d$. One has $||\vec{F}(\vec{r_{1}}-\vec{r}_{2})||>\gamma/d^{2}$.
As to $\mathcal{\vec{F}\mathrm{(\mathit{\vec{r}_{i}\mathrm{)}}}}$,
using integration by part, it can be rewritten as
$\mathcal{\vec{F}}(\vec{r}_{i})\equiv-\int_{\mathbb{R}^{3}}\dd^{3}{r_{3}}\,U(\vec{r_{i}}-\vec{r}_{3})\,{\partial}n(\vec{r}_{3};t)/{\partial\vec{r}_{3}}$,
where we used the fact that the potential $U(\vec{r})\rightarrow0$
for $r\rightarrow\infty$ and $n(\vec{r,}t)\rightarrow n=N/V$ for
$r\rightarrow\infty$. Clearly, $\mathcal{\vec{F}}(\vec{r}_{1})$
vanishes for homogeneous systems. The integrand in $\mathcal{\vec{F}}(\vec{r}_{i})$
is vanishingly small in every part of the integration domain where
the gradient of the local number density, $||{\partial}n(\vec{r};t)/{\partial\vec{r}}||$,
is vanishingly small. Let us call $L_{H}$ the typical length on which
$n(\vec{r};t)$ varies noticeably. Thus, the volume of integration
in which the integrand does not vanish is of order $L_{H}^{3}$. Consequently,
the order of magnitude of the integral defining $||\mathcal{\vec{F}}(\vec{r}_{i})||$
is $({\gamma}/{L_{H}})({n}/{L_{H}})L_{H}^{3}=\gamma nL_{H}$.
Hence, $K_{4}$ is negligible with respect to $\mathcal{L}$ if ${\gamma}/{d^{2}}>\gamma nL_{H}$.
This inequality transforms into $\mathit{\Gamma^{\mathrm{2}}<\mathrm{{\delta}/{\mathit{L_{H}}}\ll1}}$
which is compatible with the physical conditions formulated in Section~\ref{sc2}.

The free motion term $\big[\hat L_{1}^{0}+\hat L_{2}^{0}\big]g_{2}(\bold{1},\bold{2};t)$
of equation~(\ref{eq:6}) is also negligible compared to the term
$\mathcal{L}$. The latter is proportional to the force between particles
1 and 2 which is of order $\Gamma^{-\mathrm{2}}$. Indeed, one has
$||\vec{F}(\vec{r_{1}}-\vec{r}_{2})||>{\left|\gamma\right|}/{d^{2}}=\mathit{\Gamma^{-\mathrm{2}}\left|\gamma\right|n^{\mathrm{2/3}}}$
while the free motion operator $\big[\hat L_{1}^{0}+\hat L_{2}^{0}\big]$ is
independent of $\mathit{\Gamma}$.

With these arguments, what remains in first approximation from Equation~(\ref{eq:6}) is
Equation~(\ref{eq:2.7}).

\subsection{Establishing the kinetic equation~(\ref{eq:15-2})}
\label{appendixa2}

Equation~(\ref{eq:2.7}) is solved, as usual, by adding the solution of
the homogeneous part of this equation to the convolution of the propagator
of the homogeneous equation with the source term $\hat L'_{12}f(\bold{1};t)\,f(\bold{2};t)$.
Using the Fourier-transform with respect to the velocities and some
simple algebra, one gets 
\begin{align}
\tilde{g}_{2}(\vec{r}_{1},\vec{\zeta}_{1},\vec{r}_{2},\vec{\zeta}_{2};t) & =\tilde{U}(t)\,\tilde{g}_{2}(\vec{r}_{1},\vec{\zeta}_{1},\vec{r}_{2},\vec{\zeta}_{2};0)\nonumber \\
 & +\frac{\partial}{\partial\alpha}\int_{0}^{t}\frac{\dd\tau}{\tau}\,\,\tilde{U}(\alpha\tau)\,\,\tilde{f}(\vec{r}_{1},\vec{\zeta}_{1};t-\tau)\,\tilde{f}(\vec{r}_{2},\vec{\zeta}_{2};t-\tau)\mid_{\alpha=1}\label{eq:13-1}
\end{align}
where $\vec{\zeta_{1}}$ and $\vec{\zeta_{2}}$ are the Fourier variables
associated to the velocities $\vec{v_{1}}$ and $\vec{v_{2}}$ and
where 
\begin{equation}
\tilde{U}(t)=\exp[(-\frac{i}{m}\vec{F}_{12}\cdot(\vec{\zeta}_{1}-\vec{\zeta}_{2}))t]
\end{equation}
is the Fourier-transform of the propagator of the homogeneous part
of equation~(\ref{eq:2.7}).

In a first step let us assume vanishing initial correlation,
$g_{2}(\vec{r}_{1},\vec{v}_{1};\vec{r}_{2},\vec{v}_{2};0)=0$. Taking
the inverse Fourier transform of the expression in Eq.~(\ref{eq:13-1}), and plugging
the resulting formula for $g_{2}(\bold{1},\bold{2};t)$ in Eq.~(\ref{eq:3})
of $I_{1}$, one obtains, after a permutation of integrals:
\begin{align}
I_{1} & =\,\,\frac{\partial}{\partial\alpha}\int \dd^{3}v_{2}\int\frac{\dd^{3}\zeta_{1}
\dd^{3}\zeta_{2}}{(2\pi)^{6}}\,\,e^{i\vec{\zeta}_{1}\cdot\vec{v}_{1}+i\vec{\zeta}_{2}\cdot\vec{v}_{2}}
\nonumber \\
 &\times \int_{0}^{t}\frac{\dd\tau}{\tau}\int_{S_{1}}\dd^{3}r_{2}\,\frac{(-i)}{m}\vec{F}_{12}
\cdot(\vec{\zeta}_{1}-\vec{\zeta}_{2})\,e^{-\frac{i\alpha}{m}\vec{F}_{12}\cdot(\vec{\zeta}_{1}
-\vec{\zeta}_{2})\tau}\tilde{f}(\vec{r}_{1},\vec{\zeta}_{1};t-\tau)\,\tilde{f}(\vec{r}_{2},\vec{\zeta}_{2};t-\tau)\mid_{\alpha=1}.
\label{eq:17.1}
\end{align}
Since $d\ll L_{H}$, one can approximate $\tilde{f}(\vec{r}_{2},\vec{\zeta}_{2};t-\tau)$
by its value at $\vec{r_{2}}=\vec{r}_{1}$ and extract it from the
integral over $\vec{r_{2}}$ in the ball $S_{1}$. With some algebra,
Equation~(\ref{eq:17.1}) becomes:
\begin{equation}
I_{1}\approx\,\,-\frac{\partial^{2}}{\partial\alpha^{2}}\int\frac{\dd^{3}\zeta_{1}}{(2\pi)^{3}}\,\,
e^{i\vec{\zeta}_{1}\cdot\vec{v}_{1}}\int_{0}^{t}\frac{\dd\tau}{\tau^{2}}\,\,\tilde{f}(\vec{r}_{1},
\vec{\zeta}_{1};t-\tau)\,\,n(\vec{r}_{1};t-\tau)\thinspace\thinspace J\mid_{\alpha=1},
\label{eq:15-1}
\end{equation}
with 
\begin{equation}
J\equiv\int_{\mathcal{\mathcal{\mathrm{\mathit{S_{\mathrm{1}}}}}}}\dd^{3}r\,\,\,
\left(e^{-\frac{i\alpha}{m}\vec{F}(\vec{r})\cdot\vec{\zeta}_{1}\tau}-1\right).
\label{eq:51}
\end{equation}
After a change of variable $\vec{r}\rightarrow\vec{F}\vec{(r)}$ and
passing to spherical coordinates, $J$ transforms into 
\begin{equation}
J=-2\pi\left(\frac{\left|\gamma\right|\zeta_{1}\alpha\tau}{m}\right)^{3/2}
\left(\frac{2}{3}(z_{m})^{-3/2}-\int_{z_{m}}^{\infty}\dd z\,\,z^{-7/2}\sin z\right)\,\,,
\label{eq:50}
\end{equation}
where $z_{m}\equiv{\left|\gamma\right|\alpha\tau\zeta_{1}}/{d^{2}m}$.
The above integral is an incomplete Sine-integral function whose convergent power-series
expansion in $z_{m}$ (see \cite{key-17}) yields
\begin{eqnarray}
J & = & 2\pi d^{3}\thinspace\left[\frac{4}{15}\sqrt{2\pi}\left(\frac{\left|\gamma\right|\alpha\tau\zeta_{1}}{d^{2}m}\right)^{3/2}-\frac{1}{3}\left(\frac{\left|\gamma\right|\alpha\tau\zeta_{1}}{d^{2}m}\right)^{2}\right.
\nonumber\\
 & & \hspace{13mm}\left.+\frac{1}{300}\left(\frac{\left|\gamma\right|\alpha\tau\zeta_{1}}{d^{2}m}\right)^{4}+\frac{1}{3740}\left(\frac{\left|\gamma\right|\alpha\tau\zeta_{1}}{d^{2}m}\right)^{6}+\cdots\right].
\label{eq:18-1}
\end{eqnarray}
The expression~(\ref{eq:8}) of $I_{1}$ is obtained by only keeping the first term in the above series. 

Indeed, coherently with our short-time assumption, we suppose $z_{m}\ll1$
and retain only the first term of the series. The upper boundary $t$
of the time integral in equation~(\ref{eq:15-1}) must, hence, be
such that ${\left|\gamma\right|\alpha t\zeta_{1}}/{d^{2}m}\ll1$.
More explicitly, let us replace $\zeta_{1}$ by the inverse of an
average velocity $v_{av}$ and put $\alpha=1$. This transforms the
previous inequality into ${\left|\gamma\right|}/{d^{2}m}t\ll v_{av}$.
In other words, the time $t$ must be such that the velocity increment
$\Delta v={\left|\gamma\right|}/{d^{2}m}t$ acquired by particle
$1$ during time $t$ under the force of another particle at the surface
of $S_{1}$, satisfies $\Delta v\ll v_{av}$.
Finally, introducing the first term of series~(\ref{eq:18-1}) into
Eq.~(\ref{eq:15-1}), implies Eq.~(\ref{eq:8}) which,
in turn, leads to the kinetic equation~(\ref{eq:15-2}) in Section~\ref{sc2}.
However, the other terms of the series~\ref{eq:18-1} would introduce
corrections to the kinetic equation that are in integer powers of
the velocity Laplacian.\\

We are now ready to discuss the question raised in Section~\ref{sc2} (after
equation~(\ref{eq:60-1})) about the convergence of the integral
$I_{1}$. In its form~(\ref{eq:15-1}), the only place where the diverging 
force $\vec{F}(\vec{r})$ appears is the integral $J$ given by equation~(\ref{eq:51}).
As seen from its result~(\ref{eq:18-1}), $J$ converges. This comes
from the fact that the force appears only as a phase factor in the
expression~(\ref{eq:51}) of $J$.

Assuming now non-vanishing correlation at the initial
time, it is quite simple to show that the contribution to $I_{1}$
due to the first term in the right hand side of Eq.~(\ref{eq:13-1})
is given by:
\begin{equation}
-\frac{\partial}{\partial\vec{v_{1}}}.\frac{1}{m}\int_{S_{1}}d^{3}r_{2}\int \dd^{3}v_{2}\vec{F}(\vec{r}_{1}-\vec{r}_{2})
g_{2}(\vec{r}_{1},\vec{v}_{1}-\frac{t}{m}\vec{F}(\vec{r}_{1}-\vec{r}_{2});\vec{r}_{2},\vec{v}_{2}+\frac{t}{m}\vec{F}(\vec{r}_{1}-\vec{r}_{2});t=0).
\end{equation}
For non-vanishing initial correlations this term should be added to
the right hand side of Eqs.~(\ref{eq:15-2}) and~(\ref{eq:22}).
Obviously, it produces a source term in the kinetic equation. In the
solution of the latter equation, it will only add a convolution in
velocity space between the Fourier transform of the Mittag-Leffler
function $E_{3/2}(-A\,\,\zeta^{3/2}\,\,t^{3/2})$ (see Appendix B
below) and the above term. Now, a theorem~\cite{key-7-1-1} guarantees
that, for most functional forms of the initial correlation, this convolution
will lead to a contribution to the velocity distribution that possesses
an algebraic tail in $1/v^{5/2}$. Moreover, this contribution of the memory of the correlations
in the initial condition is expected to be a transient that rapidly decays with time~\cite{key-10-2}.

\section{Solution of the kinetic equation for a homogeneous state}
\label{appendixb}

A Fourier transform with respect to $\vec{v}$ and a Laplace transform
with respect to $t$ of equation~(\ref{eq:22}) give:
\begin{equation}
\hat{\tilde{\varphi}}(\vec{\zeta};w)=\frac{w^{1/2}\,\,\,\tilde{\varphi}(\vec{\zeta};0)}
{w^{3/2}\,\,\,+\,\,\,A\,|\vec{\zeta}|^{3/2}},
\label{eq:23-1}
\end{equation}
where $\tilde{\varphi}(\vec{\zeta};0)$ is the Fourier transform of
$\varphi(\vec{v},t)$ at $t=0$, $\hat{\tilde{\varphi}}(\vec{\zeta};w)$
is the Fourier-Laplace transform of $\varphi(\vec{v};t)$ and
$$
A=\frac{n\sqrt{\pi}}{5}\left(\frac{2\pi\left|\gamma\right|}{m}\right)^{3/2}.
$$
The inverse Laplace transform of $\hat{\tilde{\varphi}}(\vec{\zeta};w)$
is taken by first expanding equation~(\ref{eq:23-1}) in powers of
$A\,|\vec{\zeta}|^{3/2}$ and, then, integrating the series term by
term. This leads to the exact solution of equation~(\ref{eq:22}):
\begin{equation}
\varphi(\vec{v};t)=\int\frac{d^{3}\zeta}{(2\pi)^{3}}e^{i\vec{\zeta}\cdot\vec{v}}\,\,
\tilde{\varphi}(\vec{\zeta};0)\,\,E_{3/2}(-A\,\,|\vec{\zeta|}\,\,t^{3/2}),
\label{eq:24-1}
\end{equation}
where 
\begin{equation}
E_{\mu}(u)=\sum_{k=0}^{\infty}\frac{u^{k}}{\Gamma(\mu k+1)},
\end{equation}
is the Mittag-Leffler function of parameter $\mu$~\cite{key-13}.
We now show that the tail of this distribution is the same as that
of a Lévy-$3/2$ distribution, that is, it behaves has $1/v^{5/2}$ for large values of $v$.
Indeed, for an arbitrary fixed time $t$, the inequality $A\,\,|\zeta|^{3/2}\,\,t^{3/2}\ll1$
implies small values of $\left|\zeta\right|$ and, consequently, large
values of the velocity that correspond to the tail of the distribution.
In order to estimate the behavior of that tail we can, thus, keep only the two first terms in the 
series above and safely make the following approximation:
\begin{equation}
E_{3/2}(-A\,|\zeta|^{3/2}\,t^{3/2})\simeq e^{-c\,(\,|\zeta|\,t\,)^{3/2}},
\label{eq:26-1}
\end{equation}
with $c=({4n}/{15})({2\pi\left|\gamma\right|}/{m})^{3/2}$.
Finally, we get 
\begin{equation}
\varphi(\vec{v};t)\simeq\int\frac{d^{3}\zeta}{(2\pi)^{3}}e^{i\vec{\zeta}\cdot
\vec{v}}\,\,\tilde{\varphi}(\vec{\zeta};0)\,\,e^{-c\,(\,|\zeta|\,t\,)^{3/2}},
\end{equation}
equivalent to the velocity convolution of the initial velocity distribution
and a symmetric Lévy-$3/2$ distribution~\cite{key-2}:
\begin{equation}
\varphi(\vec{v};t)\simeq\int \dd^{3}{u}\thinspace\thinspace\varphi(\vec{u};0)
\thinspace\thinspace L_{3/2}(\vec{v}-\vec{u},\thinspace ct^{3/2}).
\end{equation}
Using a theorem in Ref.~\cite{key-7-1-1} one then shows that
for any $\varphi(\vec{v},0)$ with finite second moments, this approximation
as well as the exact solution~(\ref{eq:24-1}) have a long tail in
$1/v^{5/2}$ where $v$ is any component of the
velocity vector $\vec{v}$.

\end{document}